\documentclass{ws-ijqi}

\usepackage{amsfonts}
\usepackage{amssymb}
\usepackage{mathrsfs}
\usepackage{theorem}

\usepackage{ifpdf}\ifpdf
\usepackage{graphicx}
\fi

\newcommand{\ket}[1]{\ensuremath{\vert#1\rangle}}
\newcommand{\bra}[1]{\ensuremath{\langle #1\vert}}

\newcommand{\kb}[2]{\ensuremath{\vert #1 \rangle \langle #2 \vert}}

\def\unity{\mbox{\small 1} \!\! \mbox{1}}

\def\tr{\mathrm{tr}}

\title{HOW TO EXPLOIT LOCAL INFORMATION \\ WHEN DISTILLING ENTANGLEMENT}

\author{EARL T. CAMPBELL}
\address{University  College London, Department of Physics and Astronomy}

\begin{document}

\maketitle

\begin{abstract}
Distillation protocols enable generation of high quality entanglement even in the presence of noise.  Existing protocols ignore the presence of local information in mixed states produced from some noise sources such as photon loss, amplitude damping or thermalization. We propose new protocols that exploit local information in mixed states.  Our protocols converge to higher fidelities in fewer rounds, and when local information is significant one of our protocols consistently improves yields by 10 fold or more.   We demonstrate that our protocols can be compacted into an entanglement-pumping scheme, allowing quantum computation in distributed systems with a few qubits per location.
\end{abstract}

\section{Introduction}

Maximally entangled pairs of qubits are a powerful resource for a plethora of computational tasks that rely on quantum correlations, from quantum cryptography to quantum computation.  Though useful, entanglement is a delicate property that is difficult to produce and store.   Imperfect experimental conditions and noise from the environment produces statistical ensembles of quantum states that possess non-maximal entanglement.  Fortunately for our aspirations to develop quantum technologies, entanglement can be distilled from many imperfect \textit{base pairs} into a smaller number of more entangled pairs\footnote{Throughout we only consider two qubit mixed states.  As is common in the literature, we use pair to refer to two qubit states.  We reserve the phrase base pair for the entanglement resource obtained before any distillation has been performed.}.  

Whilst the literature on entanglement distillation is vast, it almost exclusively tailors to distillation of mixed states that are ensembles of maximally entangled states\cite{BBPSSW01a,BDSW01a,DB01a,JTSL02a,C01a}.  Such mixed states are said to be \textit{locally maximally mixed}, as local measurements on one qubit have completely random outcomes.   For all other mixed states there exist local measurements with non-zero expectation value, and so we know some \textit{local information} about the mixed state.  Some protocols propose to distill base pairs by destroying this local information in a process called \textit{twirling}\cite{BBPSSW01a},  and other protocols simply ignore the local information\cite{DEJMPS01a}.  But surely knowledge is power and no good can come from consigning local information to the flames.   This intuition is the inspiration for this study of how we can exploit local information to design more streamlined distillation protocols.  

After describing our protocols, we consider noise models grounded in experimental considerations, and compare the performance of various protocols at distilling these base pairs.  Any qubit embodied in the non-degenerate energy levels of a physical system will undergo amplitude damping and thermalization with the environment, which leads to a preference for the ground state and gives us local information.    Also, the process of producing an entangled base pair can inherently produce local information.  For example, the effect of photon loss in certain optical based procedures for entanglement generation\cite{CCFZ01a}, where the mixed state contains an incoherent component of the doubly emitting state $\ket{1,1}$.  We will analyze different approaches to distilling (a) combined photon loss and depolarizing noise; and (b) amplitude damping noise for a range of different relaxation bases.

Our protocols are best understood as building blocks.  Having applied one round of a protocol, higher fidelities are achieved by repetition, which can be achieved in a variety of ways\cite{DBCZ01a,DB01a,hartmann06,vanMeter08}: e.g. by symmetrically always pairing up identical noisy Bell pairs, or by continuously pumping with the same base pair.  The problem of identifying the best approach to repetition can be viewed as a self-contained problem, which has been called the scheduling problem\cite{vanMeter08}.  We calculate the yield when performing distillation between identical pairs, so-called symmetric scheduling. Here we find that one of our protocols consistently attains higher yields --- roughly 10 fold or greater --- in the strong local information regime.  

Since we are primarily concerned with improvements on the level of the building blocks, we will not give detailed comparisons of different approaches to scheduling.  However, it is important that our protocols can be adapted for entanglement pumping schedules that conserve spatial resources at the cost of temporal resources.  In section~\ref{sec:pumpedZinf} we extend our protocols to enable entanglement pumping, where the pairs are non-identical. We will discuss the application of entanglement pumping to distributed quantum computers with a few qubits at each local site\cite{DB01a,JTSL02a,C01a}.  Our pumped protocols demonstrate that \textit{all} possible entangled base pairs can enable quantum computing with only two or three qubits per location, a result previously known for only locally maximally mixed states. 

Before embarking on a description of our protocols, it is first necessary to review some of the existing approaches to entanglement distillation, and also to introduce some technical tools for understanding the nature of local information and its relationship to entanglement theory and local filtering theory. 

\subsection{The prior art of entanglement distillation}

A density matrix is \textit{locally maximally mixed} when all local observables have an expectation value of zero, otherwise the density matrix contains local information.  It is well known that locally maximally mixed (LoMM) states can always be locally rotated so the density matrix is diagonal in the basis of Bell states\cite{VDD01a}.  These states are entangled if and only if they have a maximally entangled fraction, $F_{\mathrm{max}}$, greater than $1/2$, where:
\begin{equation}
 F_{\mathrm{max}} = _{\mathrm{max}: \ket{\psi}} \bra{\psi}\rho\ket{\psi},
\end{equation}
which is maximized over all maximally entangled $\ket{\psi}$.  Bennett \textit{et al.} proposed the first protocol for distilling these base pairs\cite{BBPSSW01a,BDSW01a}, which was later improved by Deustch \textit{et al.}\cite{DEJMPS01a}.  The improved protocol is often called DEJMPS after the initials of its originators, and a single round proceeds as follows:
\begin{enumerate}
 \item  take two copies of $\rho$ and locally rotate\footnote{This local rotation is not explicitly prescribed in Ref.~\refcite{DEJMPS01a}. We have found that fidelity is greater when we diagonalize the 3 by 3 submatrix of the  $R_{i,j}$ matrix for $i,j=1,2,3$ (see Eq.~(\ref{eqn:Rmatrix})).  We also always apply local operations that interchange the Bell state components to maximize fidelity.  These slight improvements on the original protocol have a modest effect on increasing fidelity and yield.} such that the largest maximally entangled state is mapped to a particular Bell state, $\ket{\psi} \rightarrow \ket{\Phi^{+}}=(\ket{00}+\ket{11})/\sqrt{2}$;
 \item  perform local bilateral control-not gates, $CX_{A1}^{A2}CX_{B1}^{B2}$;
 \item  measure qubits $A1$ and $B1$ in the computational basis;
 \item  we have succeeded and keep the final state if we measure $\ket{0}_{A1}\ket{0}_{B1}$ or $\ket{1}_{A1}\ket{1}_{B1}$, otherwise we must restart with a fresh pair of $\rho$.
\end{enumerate}
For completeness, we define the familiar control-not gate:
\begin{equation} 
	CX_{x1}^{x2}= \unity_{x1}  \otimes \kb{0}{0}_{x2}  +  X_{x1} \otimes \kb{1}{1}_{x2},
\end{equation}
with $x_{2}$ as control and $x_{1}$ as target.  Throughout, we label $A1$ and $B1$ as the qubits sharing one copy of $\rho$ and $A2$ and $B2$ as sharing a second copy of  $\rho$, with $A$ and $B$ denoting the location of the qubits.  After a successful round, the fidelity,  w.r.t $\ket{\phi^{+}}$, and success probability of DEJMPS depend on only the diagonal elements of $\rho$ in the Bell basis.  If the initial fidelity exceeds $1/2$, then a successful round produces a final state with an increased fidelity.  To achieve higher fidelities DEJMPS must be repeated, with different approaches to repetition discussed in sections~\ref{sec:YIELD} and~\ref{sec:pumpedZinf}.   

DEJMPS can also be used to distill mixed states that are not of LoMM form, but its performance has no dependence on terms that are off-diagonal in the Bell basis.  As such, DEJMPS does not utilize any local information that may be contained in these off-diagonal terms.  When applying DEJMPS to non-LoMM states, it still requires $ F_{\mathrm{max}} > 1/2$ for the protocol improve fidelity towards unity.  However, there are many entangled states that this protocol fails to distill because  $F_{\mathrm{max}} \leq 1/2$.  In these instances, ignoring the off-diagonal terms effectively destroys the entanglement along with the local information.  

Though the literature has focused on LoMM states, a paper by the Horodecki family confronts the problem of non-LoMM states and proves that all entangled states can be distilled\cite{HHH01a}.  In short, the Horodeckis found a method of taking any entangled state and probabilistically producing a state with  $F_{\mathrm{max}}>1/2$, hence enabling DEJMPS to be applicable.   The Horodeckis use the idea of \textit{local filtering}, where a local Positive Operator Valued Measurement (POVM) is performed and we post-select on one outcome.   Formally, we denote a local filter on qubit $A$ by $f_{A}$, which can be any one qubit matrix with singular values of magnitude less than one\footnote{This must be true because otherwise there will exist density matrices for which the corresponding POVM measurement has a success probability of greater than 1.}.  The post-filtering state is:
\begin{equation}
 \rho_{\mathrm{HD}} = \frac{(f_{A} \otimes \unity_{B}) \rho (f_{A} \otimes \unity_{B})^{\dagger}}{\tr ( (f_{A} \otimes \unity_{B}) \rho (f_{A} \otimes \unity_{B})^{\dagger} )  },
\end{equation}
where the denominator is the probability of successfully applying the local filter, which we denote $P_{\mathrm{HD}}^{\mathrm{filter}}$.   We will not focus on the communication cost of protocols, but note that every local filtering operation requires one classical bit  of information to communicate success.  The Horodeckis have an elegant proof that for all entangled states there exists a local filter, which we call the Horodecki filter, that boosts $F_{\mathrm{max}}$ above the 1/2 threshold.   However, in the very same paper the Horodeckis identified a class of states for which the Horodecki filter works but is very inefficient.  For this class of states they also identified a different filtering operation that was superior to the Horodecki filter.   Consequently, we are left with the open question ``is there a systematic and efficient procedure for exploiting local information to distill entanglement?".   

\subsection{Local information, local filters and concurrence}

Here we review some results on the interplay between local information, local filtering and entanglement.  We begin by noting that local information of a density matrix is most apparent in its Hilbert-Schmidt decomposition:
\begin{equation}
\label{eqn:Rmatrix}
 \rho = \frac{1}{4}\left(  \sum_{i,j = 0}^{3} R_{i,j} \sigma_{i}  \otimes   \sigma_{j} \right) 
 \end{equation}
where $\sigma_{0,1,2,3}$ are the Pauli matrices $\unity, X, Y, Z$ and $\tr( \rho )=1$ entails $R_{0,0}=1$.  If we measure an observable $\sigma_{i} \otimes \sigma_{j}$, then the expectation value of the measurement outcome is $\tr ( \rho (\sigma_{i} \otimes \sigma_{j})) =R_{i,j}$. When measuring an observable local to qubit $A$, all the relevant information is encapsulated by $R_{A}=(R_{1,0}, R_{2,0}, R_{3,0})$, and similarly local measurements on qubit B have expectation values predicted by $R_{B}=(R_{0,1}, R_{0,2}, R_{0,3})$.  Representing this information as vectors $R_{A}$ and $R_{B}$ is the Bloch sphere representation of each qubit.  Hence, these six numbers describe all the local information, and when they all vanish the state has no local information and is a LoMM state.  All other entries of the matrix, $R_{i, j}$ ($i \neq 0$, $j \neq 0$), describe correlations between the two qubits.  One advantage of the $R$ matrix representation is the simplicity of how it transforms under local unitaries.  Applying a unitary $U_{A}$ to qubit $A$ is equivalent to a rotation in the Bloch sphere, and describing this rotation by a proper orthogonal matrix $O_{A}$, it follows that $R_{A} \rightarrow O_{A} R_{A} $.  Similarly, a local unitary $U_{B}$ has an associated $O_{B}$, such that the unitary effects the transformation $R_{B}\rightarrow O_{B}R_{B}$.  When the joint system is rotated by $U_{A} \otimes U_{B}$, the whole $R$ matrix transforms as:
\begin{equation}
	R \rightarrow_{U_{A} \otimes U_{B}}   
	\left( \begin{array}{cc}
		1 & 0 \\
		0 & O_{A}	\end{array} \right)	  R 	\left( \begin{array}{cc}
		1 & 0 \\
		0 & O_{B}	\end{array} \right)^{T} .
\end{equation}
Verstraete \textit{et al.}\cite{VDD01a} studied how local filters transform the $R$ matrix, and found that local filters caused Lorentz boosts on $R$.  They then showed that all physical states can be locally filtered into one of two canonical forms, one of which is LoMM:
\begin{eqnarray}
\label{RmatrixLoMM}
	R_{\mathrm{LoMM}} 	 =  \frac{1}{4}  \left( \begin{array}{cccc} 
	1 & 0 & 0 & 0 \\
	0 & R_{1,1} & 0 & 0 \\
	0 & 0 & R_{2,2}  & 0 \\
	0 & 0 & 0 & R_{3,3}  \\
	 \end{array}  \right)  & ; & 
	 \begin{array}{ccc}	 
	 R_{1,1} & = & +p_{1}-p_{2}+p_{3}-p_{4} ; \\ 
	 R_{2,2} & = &  -p_{1}+p_{2}+p_{3}-p_{4} ; \\ 
	 R_{3,3} & = & +p_{1}+p_{2}-p_{3}-p_{4} ;
	 \end{array}
\end{eqnarray}
where $p_{i}$ are the probabilistic weights of 4 Bell states.  The second canonical form they discovered they called the rank deficient (RD) class, as $\rho_{RD}$ always has rank less than the maximal 4.  The canonical form of RD states is:
\begin{eqnarray}
	R_{RD} 	& = & \frac{1}{4}   \left( \begin{array}{cccc} 
	1 & 0 & 0 & c \\
	0 & a & 0 & 0 \\
	0 & 0 & -a& 0 \\
	b & 0 & 0 & 1+b-c\\
	 \end{array}  \right); 
\end{eqnarray}
where $a,b,c$ are reals.  This form is not unique as the local information can be increased or suppressed by further local filtering.  However, the local information can never be completely filtered away. Having already remarked that Horodecki filtering is not optimal, this motivates considering whether local filtering into the appropriate canonical form will perform better.  Indeed, this will prove to be a good starting point.  However, the class of rank deficient states is measure zero and so real world mixed states are unlikely to neatly fall into the rank deficient class.  On the other hand, all full rank states can be local filtered into LoMM form, but if they contain a large amount of local information then the local filters will rarely succeed (see sections~\ref{secNOISEmodels} and~\ref{sec:YIELD} for details).  We will propose an approach that circumvents these problems.

It is also instructive to consider how the entanglement of mixed states changes after local filtering.  As an intermediate step to calculating the concurrence\cite{W01a}, we find the eigenvalues of a matrix:
\begin{equation}
	M = \sqrt{  \sqrt{\rho} \tilde{\rho} \sqrt{\rho} } ,
\end{equation}
where the tilde operation is
\begin{equation}
	\tilde{\rho} = ( Y \otimes Y) \rho^{*} (Y \otimes Y) ,
\end{equation}
with the star indicating complex conjugation in the computational basis.  This matrix will have 4 eigenvalues, labeled $\lambda_{i}$ and the concurrence is the largest eigenvalue minus the other 3. It is intriguing that local filtering may change the concurrence whilst preserving the ratios of concurrence eigenvalues.  Formally, if we denote the post-filtering concurrence eigenvalues by $\lambda'_{i}$, we find $\lambda'_{i}/\lambda'_{j}=\lambda_{i}/\lambda_{j}$.  This property is useful as it entails that the $p_{i}$ of any post-filtering $R_{\mathrm{LoMM}}$ follow directly from the concurrence eigenvalues of the original $\rho$ via:
\begin{equation}
\label{eqn:Conc_LoMMform}
	p_{i} = \lambda_{i} / ( \lambda_{1} + \lambda_{4} + \lambda_{3} + \lambda_{2}    ).
\end{equation}

\section{Two Proposed Protocols}

\begin{figure*}[t]
\centering
 \includegraphics[width=350pt]{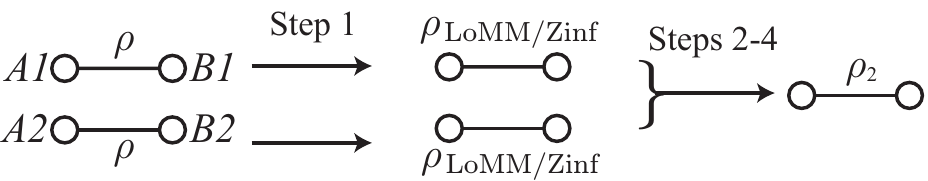}
 \caption{A schematic of the LoMM and Zinf protocols.  Both follow the same general pattern, but differ in the operation used in each step.  They both consume base pairs, $\rho$, and output a Bell diagonal mixed state $\rho_{2}$.} \label{fig:LoMMZinf}
\end{figure*}

In this section we propose two protocols for entanglement distillation, called LoMM and Zinf protocol for reasons that will be explained.  The LoMM protocol has never been explicitly proposed, but one could arrive at it by combining the single copy local filtering protocols of Refs.~\refcite{VDD01a} and~\refcite{KeLM01a} with DEMPJS.  Whilst LoMM will prove useful in some regimes, it can suffer from a pathologically small success probability when the initial mixed state contains a lot of local information.  Our more elaborate Zinf protocol is quite unique compared to previous protocols.   It is tailored to avert pathologies in the strong local information regime, and hence will outperform other known protocols in this regime.  In this section we will describe one round of these protocols (outlined in Fig.~\ref{fig:LoMMZinf}), with one round performance analyzed in section~\ref{secNOISEmodels}.    
 
\subsection{LoMM protocol}

As indicated earlier, we begin by considering distillation via locally filtering mixed states into their canonical form, as defined by Verstraete\textit{et al.}\cite{VDD01a}.  Typically, mixed states will local filter into LoMM form and so we call this the LoMM protocol.  Given a base pair, $\rho$, we apply a local filter $f_{A} \otimes f_{B}$ to bring the state into LoMM form, such that
\begin{eqnarray}
	\rho_{\mathrm{LoMM}} & = & \frac{(f_{A} \otimes f_{B})\rho(f_{A} \otimes f_{B})^{\dagger}}{\tr ( (f_{A} \otimes f_{B})\rho(f_{A} \otimes f_{B})^{\dagger}  )  },
\end{eqnarray}
which succeeds with a probability, denoted $P_{\mathrm{LoMM}}^{\mathrm{filter}}$, equal to the denominator of this expression.  	By conservation of ratios of concurrence eigenvalues under local filtering, we conclude this mixed state is:
\begin{eqnarray*}
	\rho_{\mathrm{LoMM}} & = & \big( \lambda_{1} \kb{ \Phi^{+}}{\Phi^{+}}+ \lambda_{2} \kb{ \Psi^{-}}{\Psi^{-}}   +  \lambda_{3} \kb{ \Psi^{+}}{\Psi^{+}} + \lambda_{4} \kb{ \Phi^{-}}{\Phi^{-}} \big) / \sum_{i} \lambda_{i},
\end{eqnarray*}
where $\lambda_{i}$ are the eigenvalues of the concurrence matrix of $\rho$.  Since local unitaries can interchange Bell states, we can choose $\lambda_{1}$ and $\lambda_{4}$ to be the largest and smallest $\lambda_{i}$, respectively.  After two successful attempts at local filtering, $\rho \rightarrow \rho_{\mathrm{LoMM}}$, we have two copies of $\rho_{\mathrm{LoMM}}$.   Now we have two copies of a Bell diagonal state it can be distilled in the usual fashion, and so the whole LoMM protocol is:
\begin{enumerate}
 \item  repeatedly attempt to local filter copies of $\rho$ into LoMM form until we have two copies of $\rho_{\mathrm{LoMM}}$;
 \item  perform local bilateral control-not gates, $CX_{A1}^{A2}CX_{B1}^{B2}$;
 \item  measure qubits $A1$ and $B1$ in the computational basis;
 \item  we have succeeded and keep the final state if we measure $\ket{0}_{A1}\ket{0}_{B1}$ or $\ket{1}_{A1}\ket{1}_{B1}$, otherwise we must restart with a fresh pair of $\rho_{\mathrm{LoMM}}$.
\end{enumerate}
The probability of success for steps 2-4 is:
\begin{equation}
	P_{\mathrm{LoMM}}^{\mathrm{distil}} = \left( (\lambda_{1}+\lambda_{4})^{2}  +  (\lambda_{3}+\lambda_{2})^{2} \right) / \sum_{i}\lambda_{i},
\end{equation}
and the final state is:
\begin{eqnarray}
	\rho_{2}  & \propto &(\lambda_{1}^{2}+\lambda_{4}^{2}) \kb{\Phi^{+}}{\Phi^{+}} +  2\lambda_{1}\lambda_{4} \kb{\Phi^{-}}{\Phi^{-}}   \\ \nonumber
	& + &   (\lambda_{3}^{2}+\lambda_{2}^{2}) \kb{\Psi^{+}}{\Psi^{+}} +  2\lambda_{3}\lambda_{2} \kb{\Psi^{-}}{\Psi^{-}}.
\end{eqnarray}
Recall that we had the foresight to fix $\lambda_{1}$ and $\lambda_{4}$ as the largest and smallest $\lambda_{i}$, so that the maximally entangled fraction:  
\begin{equation}
\label{OneRoundFid}
	F'_{\mathrm{max}} = \frac{  \lambda_{1}^{2}+\lambda_{4}^{2} }{  (\lambda_{1}+\lambda_{4})^{2}  +  (\lambda_{3}+\lambda_{2})^{2} } ,
\end{equation}
cannot be increased by any reordering\footnote{We reorder $\lambda_{i}$ by applying the appropriate local unitaries to $\rho_{\mathrm{LoMM}}$.} of the $\lambda_{i}$.  Note also, that entanglement increases whenever the base pair was entangled. 

Whilst the LoMM protocol certainly has it merits, in later sections we will see that $P_{\mathrm{LoMM}}^{\mathrm{filter}}$ can be extremely small.  Before turning to this analysis we will first introduce our second, more sophisticated,  protocol.

\subsection{Zinf protocol}
\label{secZinfdistil}
 
Since it is already known that we can always find local filters that bring the Hilbert-Schmidt decomposition into a form  $R_{\mathrm{LoMM}}$ or $R_{RD}$, it follows that all states can be locally filtered to some linear combination of these possibilities:
\begin{eqnarray} 
	R_{\mathrm{Zinf}} & = & \frac{1}{4} \left( \begin{array}{cccc} 
	1 & 0 & 0 & R_{3,0} \\
	0 & R_{1,1} & 0 & 0 \\
	0 & 0 & R_{2,2}& 0 \\
	R_{0,3} & 0 & 0 & R_{3,3}\\
	 \end{array}  \right),
\end{eqnarray}
where the label ``Zinf" emphasizes that the only off-diagonal contributions of $R_{\mathrm{Zinf}} $ are from the local $Z$-\textit{inf}ormation.  Our second protocol prescribes local filtering into Zinf form.  However, for any given $\rho$ there is a whole family of appropriate local filters that perform a transformation $\rho \rightarrow \rho_{\mathrm{Zinf}}$.  Any local filters that belong to this family we denote $g_{A} \otimes g_{B}$, so:
\begin{equation}
	\rho_{\mathrm{Zinf}} = \frac{(g_{A}\otimes g_{B} )\rho(g_{A}\otimes g_{B} )^{\dagger}}{\tr ( (g_{A}\otimes g_{B} )\rho(g_{A}\otimes g_{B} )^{\dagger} )}.
\end{equation}
Recall that the denominator gives the local filter's success probability, $P_{\mathrm{Zinf}}^{\mathrm{filter}}$.  If we already know that local filter $f_{A} \otimes f_{B}$ brings $\rho$ in LoMM form, then the family of $g_{A}\otimes g_{B}$ local filters is:
\begin{equation}
	g_{A} \otimes g_{B} = (z_{A} \otimes z_{B})(u_{A} \otimes u_{B})(f_{A} \otimes f_{B}),
\end{equation}
where $u_{A} \otimes u_{B}$ are local unitaries that interchange Bell states, and $z_{A} \otimes z_{B}$ are computational basis local filters such that:
\begin{equation}
	z_{i} \propto c_{i}  \kb{0}{0} + (1-c_{i}) \kb{1}{1},
\end{equation}
with real $c_{A}$ and $c_{B}$. Hence, this family of local filters has a simple description.

Given this family of local filters, we are free to choose the local filter according to any figure of merit we desire, which will vary depending on the ultimate application of the distillation protocol.  In the forthcoming analysis we select the local filter that maximizes the overall success probability:
\begin{equation}
 P_{\mathrm{Zinf}}^{\mathrm{all}}= (P_{\mathrm{Zinf}}^{\mathrm{filter}})^{2} P_{\mathrm{Zinf}}^{\mathrm{distil}}.
\end{equation}
This is prudent as it can significantly boost the success probability above that attained by the LoMM protocol.  Amazingly, we will see that whatever $\rho_{\mathrm{Zinf}}$ we choose to locally filter to, our distillation protocol will output the same final state up to an interchange of the concurrence eigenvalues.  At this juncture it is timely to explicitly expand $\rho_{\mathrm{Zinf}}$ into bra-ket notation:
\begin{eqnarray}
	\rho_{\mathrm{Zinf}} & = & p_{1} \kb{ \Phi^{+}}{\Phi^{+}} + p_{2} \kb{ \Phi^{-}}{\Phi^{-}}  +  r \left(  \kb{ \Phi^{+}}{\Phi^{-}} + \kb{ \Phi^{-}}{\Phi^{+}} \right) \\ \nonumber
	 & + & p_{3} \kb{ \Psi^{+}}{\Psi^{+}} + p_{4} \kb{ \Psi^{-}}{\Psi^{-}}  +  s \left(  \kb{ \Psi^{+}}{\Psi^{-}} + \kb{ \Psi^{-}}{\Psi^{+}} \right),
\end{eqnarray}
where $p_{i}$ follow from Eq.~(\ref{RmatrixLoMM}) and the off-diagonal terms relate to the local information, via $r=(R_{3,0}+R_{0,3})/4$ and $s=(R_{3,0}-R_{0,3})/4$.   For subsequent reference, we note that the concurrence eigenvalues of $\rho_{\mathrm{Zinf}}$ are:
\begin{eqnarray}
\label{eqnCONCeigenvalues}
	\begin{array}{rclc} 
	\lambda_{1}&=&\sqrt{A+B} ;  \\
	\lambda_{2}&=&\sqrt{C-D} ; \\
	\lambda_{3}&=&\sqrt{C+D} ;  \\
	\lambda_{4}&=&\sqrt{A-B} ;
	\end{array}
	 &   
\begin{array}{crl} 
	A & = & (p_{1}^{2}+p_{2}^{2})/2-r^{2} ; \\ 
	B & = & (p_{1}-p_{2}) \sqrt{ (p_{1}+p_{2})/2- r^{2} } ; \\
	C & = & (p_{3}^{2}+p_{4}^{2})/2-s^{2} ; \\ 
	D & = & (p_{3}-p_{4}) \sqrt{ (p_{3}+p_{4})/2- s^{2} } .
\end{array} & 
\end{eqnarray}
Once two copies of $\rho_{\mathrm{Zinf}}$ have been successfully prepared, we are ready to perform distillation.  However, since $\rho_{\mathrm{Zinf}}$ is not LoMM we have to adapt the usual approach:
\begin{enumerate}
 \item  repeatedly attempt to local filter copies of $\rho$ into Zinf form until we have two copies of $\rho_{\mathrm{Zinf}}$;
 \item  perform local bilateral control-not gates, $CX_{A1}^{A2}CX_{B1}^{B2}$;
 \item  measure qubits $A1$ and $B1$ in the computational basis;
 \item  we have succeeded and keep the final state if we measure $\ket{1}_{A1}\ket{1}_{B1}$, otherwise we must restart with a fresh pair of $\rho_{\mathrm{Zinf}}$.
\end{enumerate}
The main difference to conventional distillation is that we post-select on only the $\ket{1}_{A1}\ket{1}_{B1}$ measurement outcome, which occurs with probability:
\begin{eqnarray}
   P_{\mathrm{Zinf}}^{\mathrm{distil}} & = & \frac{1}{2} \big(  (p_{1}+p_{2})^{2} + (p_{3}+p_{4})^{2} -4(r^{2}+s^{2}) \big) , \\ \nonumber
    & = & \frac{1}{2}\big(  (\lambda_{1}+\lambda_{4})^{2}  +  (\lambda_{3}+\lambda_{2})^{2}  \big) .
\end{eqnarray}
Notice that the success probability is upper bounded by $1/2$,  because we post-select only one measurement outcome.  In contrast, both DEJMPS and LoMM post-select on two possible outcomes, e.g.  $\ket{1}_{A1}\ket{1}_{B1}$ or  $\ket{0}_{A1}\ket{0}_{B1}$, which allows them to approach unit success probability.   Although this is a disadvantage for Zinf,  this more restrictive post-selection can cope with the presence of local Z information and hence proves invaluable in distilling states with large amounts of local information.  Upon success A2 and B2 are projected into the final state:
\begin{eqnarray*}
	\rho_{2} & \propto &  ((p_{1}^{2}+p_{2}^{2})/2-r^{2}) \kb{\Phi^{+}}{\Phi^{+}} +  (p_{1}p_{2}-r^{2}) \kb{\Phi^{-}}{\Phi^{-}}   \\ \nonumber
	 &+&   ((p_{3}^{2}+p_{4}^{2})/2-s^{2}) \kb{\Psi^{+}}{\Psi^{+}} +  (p_{3}p_{4}-s^{2}) \kb{\Psi^{-}}{\Psi^{-}} ,
\end{eqnarray*}
which can be simplified using Eq.s~(\ref{eqnCONCeigenvalues}): 
\begin{eqnarray}
\label{Zinfresult}
 	\rho_{2}  &\propto &(\lambda_{1}^{2}+\lambda_{4}^{2}) \kb{\Phi^{+}}{\Phi^{+}} +  2\lambda_{1}\lambda_{4} \kb{\Phi^{-}}{\Phi^{-}}   \\ \nonumber
	& + &   (\lambda_{3}^{2}+\lambda_{2}^{2}) \kb{\Psi^{+}}{\Psi^{+}} +  2\lambda_{3}\lambda_{2} \kb{\Psi^{-}}{\Psi^{-}}.  \nonumber
\end{eqnarray}
Again we see that Zinf produces a final state that \textit{appears} to depend on only the concurrence eigenvalues, seeming to imply that Zinf always attains the same fidelities as LoMM.  However, there is a subtle difference between the fidelities achieved by these protocols.  For the LoMM protocol we applied local unitaries to $\rho_{\mathrm{LoMM}}$ ensuring that $\lambda_{1}>  \lambda_{2},\lambda_{3}\geq \lambda_{4}$.  Here, the local unitaries, $u_{A} \otimes u_{B}$, that interchange Bell states are buried in the middle of $g_{A}\otimes g_{B}$, and consequentially they can affect the filter success probability.   If we local filter to maximize $P_{\mathrm{Zinf}}^{\mathrm{all}}$, then sometimes the $\lambda_{i}$ will be optimally ordered and sometimes not.  Note that we cannot simply interchange Bell states after applying $g_{A}\otimes g_{B}$ as applying local unitaries to $\rho_{\mathrm{Zinf}}$ can change local Z information into local  X information. 

\section{One Round Performance}
\label{secNOISEmodels}

We have reviewed the DEJMPS protocol and Horodecki distillation and proposed our own protocols, LoMM and Zinf.  We now turn to a comparison of the ``one round"  performance of these protocols for two families of physical noise model.  A ``many round" analysis will follow in subsequent sections.

\begin{figure*}[t]
\centering
 \includegraphics[width=350pt]{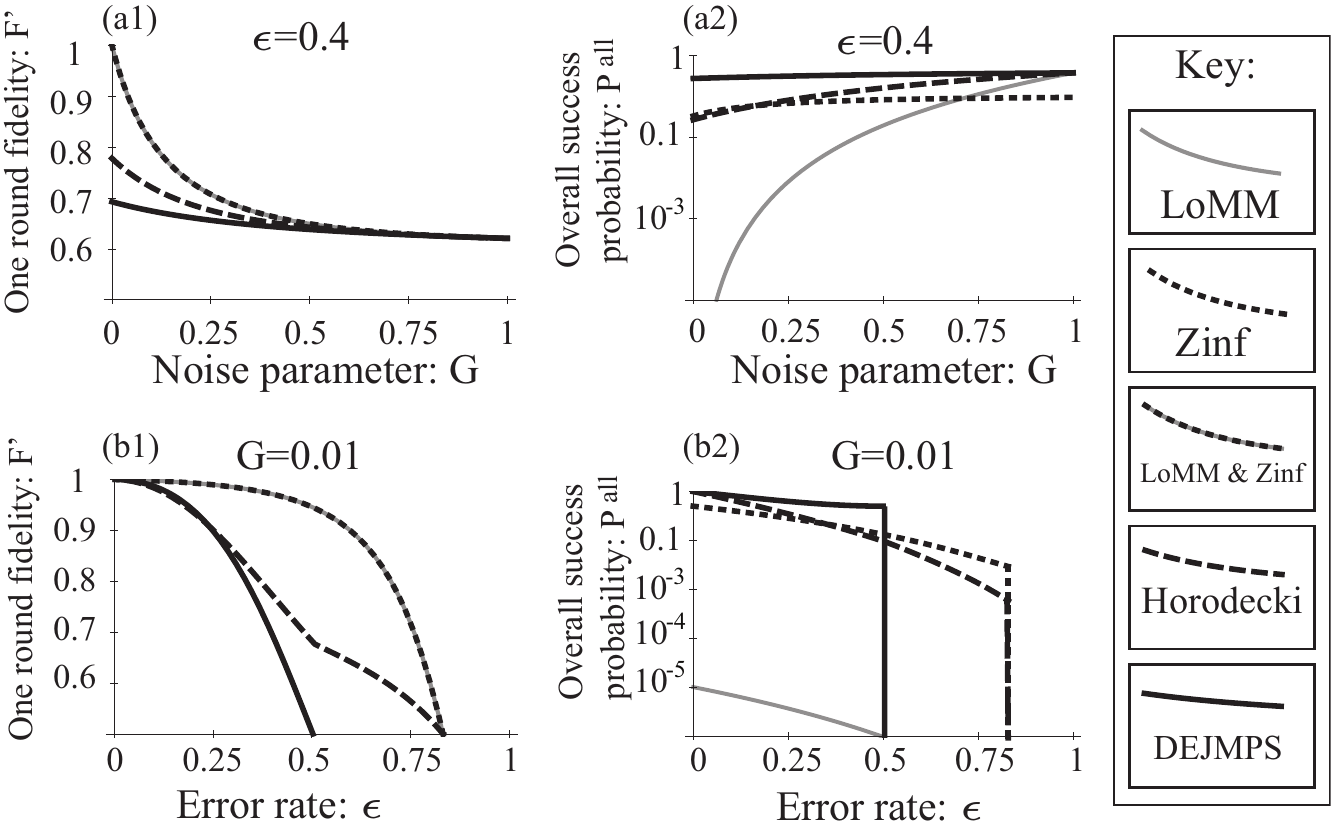}
 \caption{One round performance of distillation of photon loss noise (see Eq.~(\ref{RHOphotonloss})) for various protocols.  Plots (a1) and (a2) have the error rate $\epsilon$ fixed at 0.4 with $G$ varying. Plots (b1) and (b2) have $G$ fixed at 0.01 and $\epsilon$ varying.  The resulting fidelity after a single successful round, $F'$, is shown in the (a1) \& (b1) plots.  Note that, in the (a1) \& (b1) plots the LoMM and Zinf protocols give identical resulting fidelities, which is shown by a grey curve overlaid with black dots.   The (a2) \& (b2) plots give the overall success probability, $P^{\mathrm{all}}=(P^{\mathrm{filter}})^{2}P^{\mathrm{distil}}$, which combines the probability of local filtering twice and distilling.}
 \label{FigPhotonLossNoise}
\end{figure*}

\subsection{From photon loss to generic noise}
\label{secPhotonLoss}

Experiments have demonstrated entanglement over macroscopic separations via measurements on photons\cite{MMOYMDM1a,MMOYMM01a}.  Building on this success, a number of promising schemes for quantum computation propose projecting matter qubits into an entangled state depending on photon measurements\cite{CCFZ01a,BKPV01a,FZLGX01a,BPH01a,LBK01a,BK01a,B01a,BES01a,LBBKK01a,BBFM01a}.  However, some schemes are not robust against photon loss, which can cause the projected state to contain a large proportion of the $\ket{1,1}$ component.  Here we consider mixed states damaged by a combination of photon loss noise and depolarizing noise:
\begin{eqnarray}
\label{RHOphotonloss}
 	\rho_{\mathrm{PL}} & = & (1- \epsilon) \kb{\Psi^{+}}{\Psi^{+}} + \frac{\epsilon}{1+2G} \kb{1,1}{1,1} \\ \nonumber
	  & + &   \frac{G \epsilon }{1+2G}  \left(  \kb{0,0}{0,0}  +  \kb{\Psi^{-}}{\Psi^{-}} \right) .
\end{eqnarray}
When $G=0$, there is only photon loss noise, and the mixed state will locally filter to a rank deficient state.  When $G=1$, all noise is depolarizing and $\rho_{\mathrm{PL}}$ is LoMM without any local filtering.  For $0<G<1$ we have the intermediary case and the base pair can be locally filtered into a LoMM canonical form.      Furthermore, the base pair inherently has all local information in the Z-basis, and so Zinf can be performed without the filtering step.  Distillation of photon loss noise has been considered previously\cite{CB01a,BDSW01a}, and the best-known protocols are instances of our more general Zinf protocol.

In Fig.~(\ref{FigPhotonLossNoise}) we show the performance of four different protocols at distilling $\rho_{\mathrm{PL}}$.  With the noise rate, $\epsilon$, held constant ($\epsilon=0.4$) and $G$ varying, we see that as $G \rightarrow 1$ all protocols converge to the same output fidelity.  In this limit, we also observe that the overall success probability, $P^{\mathrm{all}}$, converges to the same value for all protocols except Zinf, which suffers an inherent $1/2$ overhead.  However, we are really interested in the strong local information regime, $G \rightarrow 0$, where photon loss dominates.  In this regime Zinf and LoMM protocols produce exceptionally high fidelity states after only a single  successful round of distillation, with Horodecki distillation and DEJMPS performing significantly worse.  Whilst Zinf and LoMM give identical outputs on success, LoMM has an overall success probability that vanishes with G as $P_{\mathrm{LoMM}}^{\mathrm{all}} \sim (1-\epsilon)^{4}G^{2} + O[G^{4}]$. 

Although our protocols achieve higher fidelities in a single round this comes at the cost of lower success probability, which can be a significant cost for LoMM.   In section~\ref{sec:YIELD}, we consider one approach to applying many rounds of distillation, and find that the yield, which depends on both fidelity and success probability, is greatest for Zinf in the regime of interest.

\begin{figure*}[t]
\centering
 \includegraphics[width=350pt]{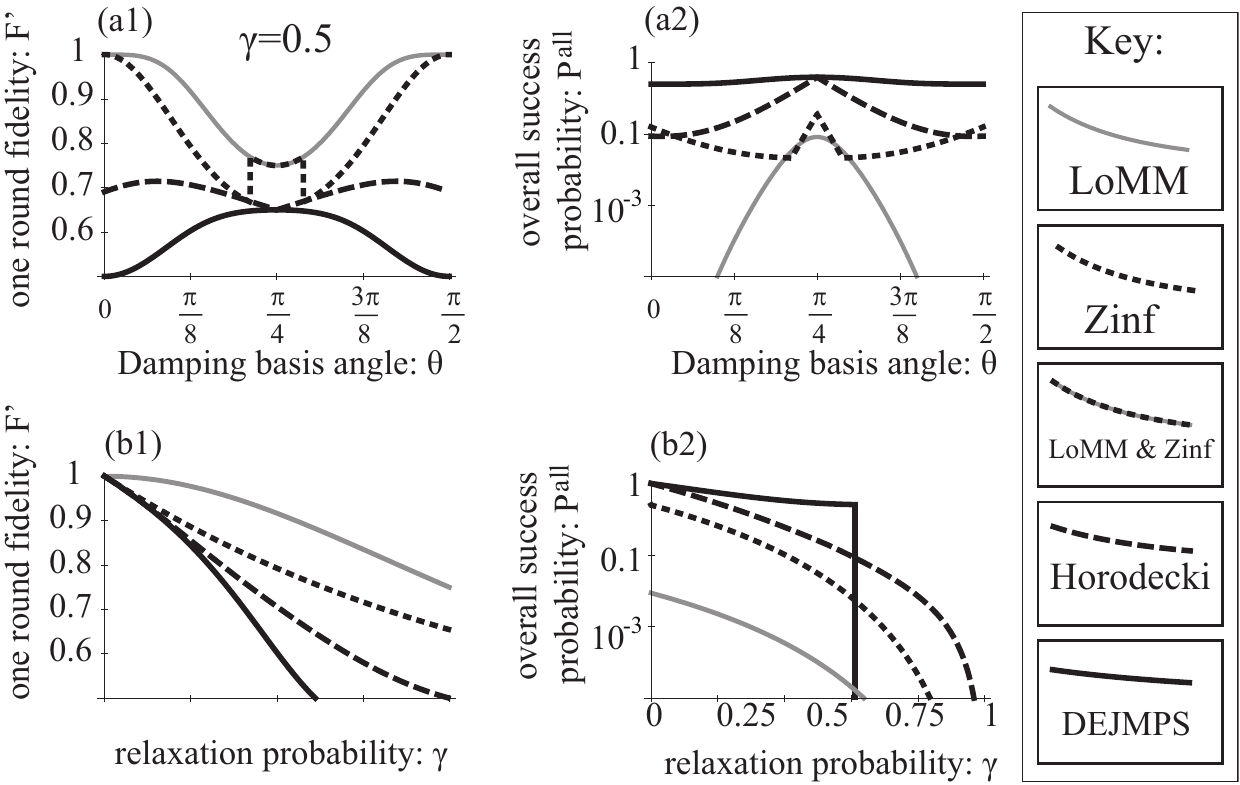}
 \caption{One round performance of distillation of photon loss noise (see Eq.~(\ref{eqnAmpDamp})) for various protocols. Plots (a1) and (a2) have the relaxation rate $\gamma$ fixed at 1/2  and vary $\theta$. Plots (b1) and (b2) have the damping angle $\theta$ fixed at $\pi/8$ and $\gamma$ varying.  The resulting fidelity after a single successful round, $F'$, is shown in the (a1) \& (b1) plots.   The (a2) \& (b2) plots give the overall success probability, $P^{\mathrm{all}}=(P^{\mathrm{filter}})^{2}P^{\mathrm{distil}}$, which combines the probability of local filtering twice and distilling. }
 \label{FigAmpDamp}
\end{figure*}

\subsection{Amplitude damping}
\label{sec:AMPdamp}

When the logical states of a qubit correspond to the excited and ground state of a physical system, it can gradually decay down to the ground state.  Such a process is known as amplitude damping, and despite its prevalence in the laboratory the present author is not aware of any studies of distilling entanglement from amplitude damped states.  Furthermore, since amplitude damping increases local information it is a natural choice as a test bed for our proposed protocols.  Hence, we consider the family of base pairs:
\begin{eqnarray}
\label{eqnAmpDamp}
	\rho_{\mathrm{AD}} & = & \sum_{i,j = 1,2} (K_{A}^{i} \otimes K_{B}^{j}) \kb{\Psi^{+}}{\Psi^{+}} (K_{A}^{i} \otimes K_{B}^{j})^{\dagger}  .
\end{eqnarray}
Here $K_{A}^{j}$ and $K_{B}^{j}$ are local Kraus operators acting on qubits $A$ and $B$, such that:
\begin{eqnarray}
 	K_{x}^{1} & = & U^{\theta}_{x} \left( \kb{0}{0}_{x}+ \sqrt{1-\gamma} \kb{1}{1}_{x}  \right) U_{x}^{-\theta} ,\\ \nonumber
	K_{x}^{2} & = & U^{\theta}_{x} \left( \sqrt{\gamma} \kb{0}{1}_{x} \right)  U^{-\theta}_{x},
\end{eqnarray}
where $\gamma$ is the relaxation probability, and the unitary $U_{x}^{\theta}$ is:
\begin{equation}
	U_{x}^{\theta}=  \cos(\theta) \unity_{x} + i \sin (\theta) Y_{x}.
\end{equation}
When $\theta=0$, the base pair $\rho_{\mathrm{AD}}$ describes computational basis amplitude damping acting on the $\ket{\Psi^{+}}$ state.  When $\theta \neq 0$ the effect of the noise channel can be interpreted as either amplitude damping in some non-computational basis, or equivalently computational basis damping that has acted on some state  $(U_{A}^{\theta} \otimes U_{B}^{\theta})\ket{\Psi^{+}}$.

Figure~(\ref{FigAmpDamp}) depicts the performance of our protocols and Horodecki distillation for a range of parameters, $\gamma$ and $\theta$.  After a successful first round, both LoMM and Zinf always produce states with a greater fidelity than DEJMPS or Horodecki distillation.  This is most pronounced when amplitude damping approaches the computational basis,  $\theta \rightarrow 0$, where both our protocols approach an output of unit fidelity.  In contrast, DEJMPS approaches a separable state in this limit\footnote{This is because of our choice of $\gamma=0.5$.   For  $\gamma < 0.5$, DEJMPS would produce an entangled state for all values of $\theta$.  Whereas  for $\gamma>0.5$ there exists a non-zero threshold value of $\theta$, below which DEJMPS produces a separable state.   When the parameters, $\gamma$ and $\theta$, are such that DEJMPS produces a separable state, the initial base pair has a maximally entangled fraction below 1/2, formally $F_{\mathrm{MAX}} \leq 1/2$.} as for $\theta=0$ and $\gamma \leq 1/2$ the initial base pair has a maximally entangled fraction below 1/2.  In the same limit, we observe the now familiar feature of the LoMM protocol having a vanishing success probability, whereas all other protocols maintain a finite success probability.  Again our protocols offer increased fidelity in exchange for reduced success probability.  In section~\ref{sec:YIELD}  we consider the overall balance of these effects, where we will find that Zinf has the greatest yield in the small $\theta$ regime, and all other protocols converge to a similar performance as $\theta \rightarrow \pi/4$.

Distillation of amplitude damping noise also presents some interesting features that we did not encounter for photon loss noise.  Most notably, the Zinf protocol exhibits a striking discontinuity as $\theta$ varies: it jumps between producing the same output as LoMM and producing a less entangled output.  This discontinuity occurs because we have specified that Zinf uses the local filter that maximizes the quantity $P_{\mathrm{Zinf}}^{\mathrm{all}}$.  The two different branches of the discontinuity correspond to local filters from different families: (a) the family of local filter that produces $\rho_{\mathrm{Zinf}}$ with the variables $\lambda_{i}$ ordered optimally (e.g. $\lambda_{1}$ largest, $\lambda_{4}$ smallest); or (b) the family of local filters with a suboptimal ordering of $\lambda_{i}$ variables.  However, the reader should note that the choice of Zinf local filter is flexible, and for some tasks it may benefit from local filters with optimally ordered $\lambda_{i}$.  

\begin{figure*}[t]
\centering
 \includegraphics[width=350pt]{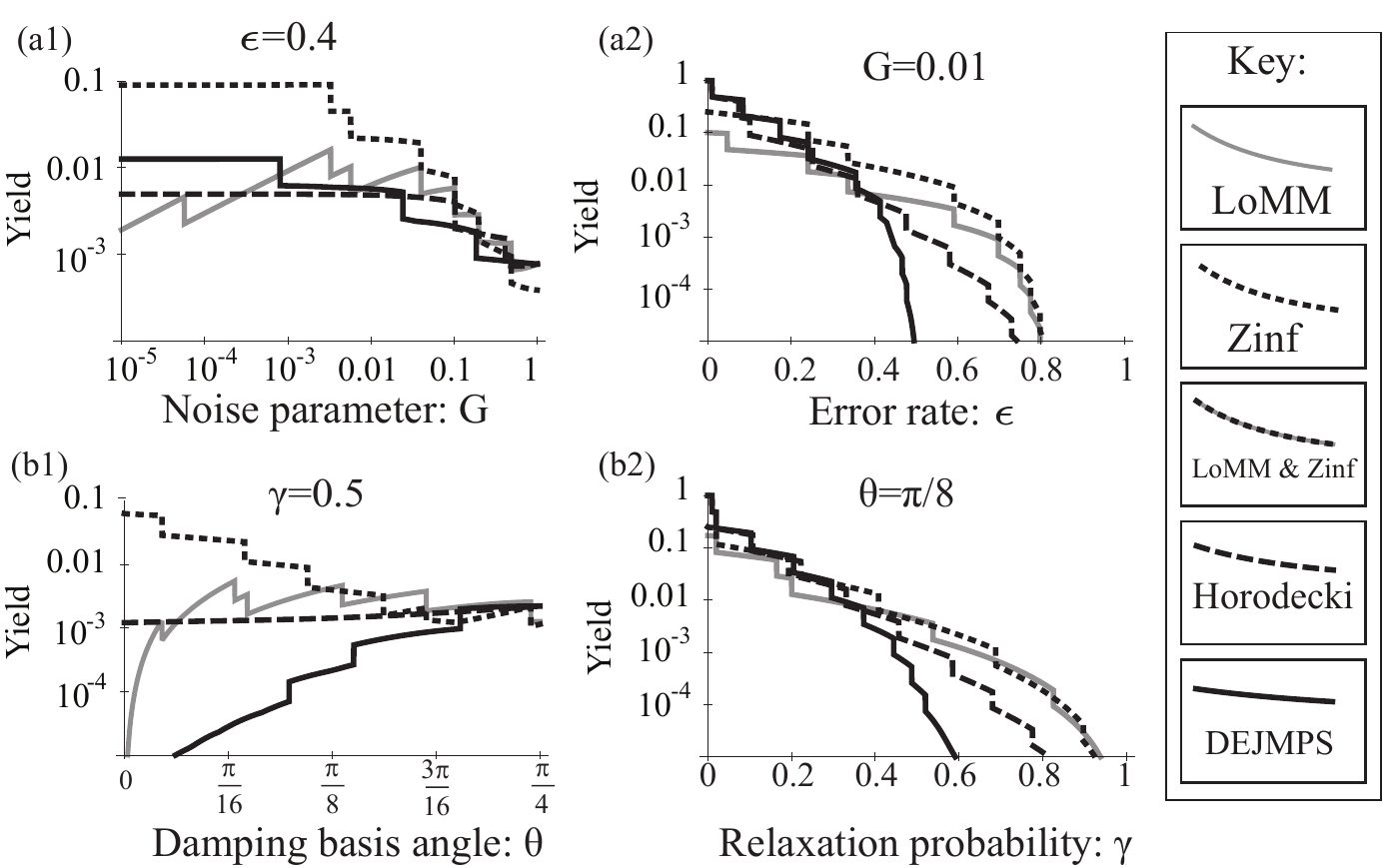}
 \caption{The yield (the average number of Bell pairs of fidelity $F=0.99$ produced per base pair consumed) for various protocols when the base pair suffers (a) photon-loss/depolarizing noise (see section~\ref{secPhotonLoss} ), and (b) amplitude damping (see section~\ref{sec:AMPdamp} ).  Local information is greatest for (a) large $\epsilon$ and small $G$, or (b) large $\gamma$ and small $\theta$.  Zinf (dotted line) is the only protocol that consistently has the largest yield in these regimes, often gaining over an order of magnitude over prior protocols. Note that the ranges of yields covered is not the same across the four figures, and in particular (a1) and (b1) only go up to a yield of 0.1.}
 \label{figYIELDSymmetric}
\end{figure*}

\section{Symmetric Scheduling}
\label{sec:YIELD}

So far we have only considered one round of entanglement distillation, which may not deliver the required fidelity.  Typically we need to repeat many rounds of distillation.  However, the optimal means of repetition depends on a number of factors, including:   the number of qubits at each location; whether the procedure forms a component in a repeater architecture\cite{DBCZ01a}; and whether memory effects are important\cite{hartmann06}.  The simplest scenario is when there are no size limits on the local Hilbert space, memory effects are neglected and we aim to produce Bell pairs between two fixed locations (rather than via several ``hops" in a repeater architecture).  This simple scenario admits a simple, symmetric, approach to repeating distillation. In symmetric scheduling, many copies of the base pairs are simultaneously stored, and simultaneously distilled to produce an output denoted $\rho_{2}$.  Pairs of $\rho_{2}$ are then distilled into an output $\rho_{3}$, and so on.  

For Zinf\footnote{Note that the Zinf filter has been chosen to maximize yield.} and LoMM, the output of the first round is a Bell diagonal mixed state that no longer contains local information.  Therefore, our protocols will only be used in the first round of symmetric scheduled approaches, with subsequent rounds performed by DEJMPS. Nevertheless, Fig.~\ref{figYIELDSymmetric} shows a marked difference in the yield for all protocols.  The yield depends on both the one round fidelity and success probability, and so the yield results should be compared alongside the one round performance.  Specifically, low yields occur whenever the one round success probability vanishes or the one round fidelity approaches 0.5 (the value for a separable Bell diagonal state).

We see that for Zinf the reduced success probability can be worth the increased fidelity.  Indeed, in strong local information regimes Zinf can produce yields 10 times greater than prior protocols.  Our results show that LoMM can sometimes match Zinf's yield in regimes of modest local information.  However, LoMM has a rapidly vanishing yield when the local information becomes very significant ( e.g. $G \rightarrow 0$ or $\theta \rightarrow 0$).  This yield crash of LoMM is due to a rapidly decreasing success probability for the local filter into LoMM form.  

In contrast, as  $\theta \rightarrow \pi / 4$ or $G \rightarrow 1$, local information decreases and all protocols converge to similar yields.  Although yields are similar in this regime, the Zinf protocol can perform slightly worse.  This is because Zinf postselects on fewer measurement outcomes, and so has a factor of 1/2 overhead.

\section{Entanglement Pumping}
\label{sec:pumpedZinf}

Although symmetric scheduling offers a rapid means of producing high fidelity Bell pairs, it requires a large Hilbert space to simultaneously store many copies of $\rho$.  This spatial cost can be very large, and in many instances the available technology may not be able to provide such a large Hilbert space.  In these scenarios, it is advantageous to perform distillation between noisy Bell pairs that are not identical.  Such approaches are asymmetric, as in each distillation attempt one participating Bell pair will have a higher fidelity than the other.  For example, the most asymmetric of approaches is entanglement pumping where one of the two pairs is always a base pair.  The advantage of entanglement pumping is that we only have to store two pairs at any one time.  However, the price of conserving Hilbert space is a slower convergence to higher fidelities, and a reduced yield.  Since entanglement pumping is an important application, in subsection~\ref{sec:pumpedZinf} we show that Zinf can be modified to this task.  In our original account of Zinf it was performed between two identical Bell pairs, and the local information cancelled out.  Here we show that the local information still cancels out provided that we pump for an even number of rounds.  By using several levels of entanglement pumping one can achieve higher fidelities.   

In subsection~\ref{sec:pumpedZinf} we give a brief overview of the minimal number of pumping levels required by prior proposals.  We compare this with our pumped-Zinf protocol, which proves to be conservative in its spatial requirements.

\subsection{Overview of few-qubit pumping schemes}

D\"{u}r and Briegel\cite{DB01a} proposed a method of using many pumping levels to enable quantum computing with only a small amount of spatial resources.   In the first pumping  level, two copies of the base pair $\rho$ are distilled by a single round of DEJMPS to produce  $\rho_{2}$.  Still in the first pumping level, we perform DEJMPS using one copy of $\rho$ and one copy of  $\rho_{2}$, which upon success produces a mixed state $\rho_{3}$.  This process is continued using one copy of $\rho$ and one copy of $\rho_{i}$, until after $n$ rounds the fidelity plateaus for $\rho_{n}$.  This resource, $\rho_{n}$, then plays the role of the base pair in the next level of pumping.  If we denote $\rho_{n}=\rho'$, then the next level of pumping proceeds in the same fashion with $\rho'$ replacing $\rho$.  After some $m$ rounds in the $2^{\mathrm{nd}}$ pumping level, we have the state $\rho'_{m}$.  This then forms the resource in the $3^{\mathrm{rd}}$ pumping level, such that $\rho''=\rho'_{m}$, and so on.

The D\"{u}r and Briegel\cite{DB01a} proposal uses $N+2$ qubits per location to support $N$ pumping levels.  Of these qubits, $N+1$ play an active role in distillation, and one qubit stores a logical qubit involved in the computation.  The overall strategy is to distill a high fidelity Bell pair within the pumping levels and then use it to implement a two-qubit unitary between two separated logical qubits.   D\"{u}r and Briegel found that two or three pumping levels, and hence four or five qubits per location, attained a high enough fidelity for fault tolerant quantum computing.

Jiang \textit{et al.}\cite{JTSL02a} proposed a slight variation on pumped-DEJMPS that, assuming perfect local operations, enables quantum computing with any entangled LoMM base pair using only two pumping levels, and hence four qubits.  Base pairs of LoMM form with a vanishingly small $4^{\mathrm{th}}$ eigenvalue, such as phase noisy states, can be distilled using only a single pumping level.  If the available Hilbert space is smaller still, then by employing measurement-based quantum computing (MBQC), the same base pairs can enable quantum computation with the use of one less qubit\cite{C01a}.   This one qubit saving is possible because the entanglement resource for MBQC, e.g. a cluster state, can be grown by probabilistic parity projections\cite{BK01a,B01a,BR02a,DR01a}.

\subsection{Pumped-Zinf}

Whilst the Hilbert space requirements are well documented for LoMM base pairs, other mixed states have been neglected.  Hence, it is informative to consider the Hilbert space requirements of applying LoMM, Zinf or Horodecki distillation to non-LoMM mixed states.  All these protocols begin by local filtering the base pair. However, most physical systems do not admit direct POVM measurements for local filtering, and so they have to be implemented indirectly.  Fortunately, it is well known that a single ancillary qubit can be used to implement any two-outcome POVM measurement.  Hence, one may expect that distillation of general mixed states may require at least one more qubit per location than required for distilling LoMM base pairs.

\begin{figure*}[t]
\centering
 \includegraphics{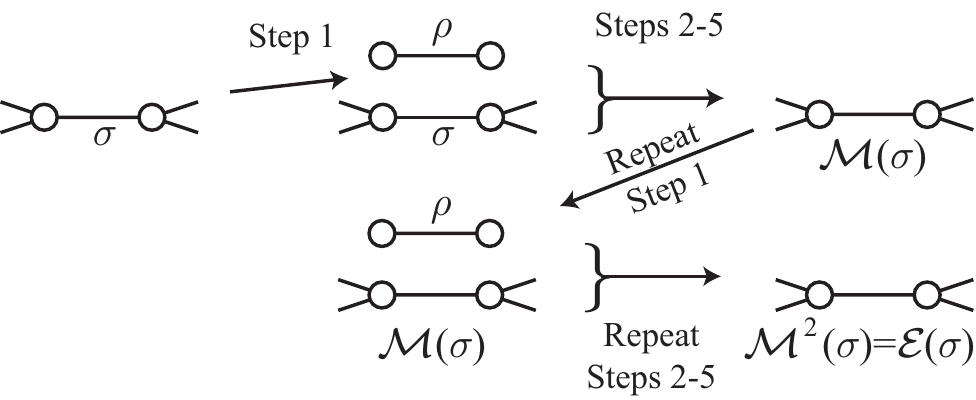}
 \caption{A schematic of pumped-Zinf (a similar pumped version of LoMM is also possible). Entanglement pumping uses base pairs, $\rho$, to increase entanglement of a state $\sigma$; e.g. via noisy parity projections, .  The state $\sigma$ may be the output of a previous round of Zinf, e.g. the state $\rho$ from Fig.~\ref{fig:LoMMZinf}.  We always perform pumped-Zinf an even number of times, consuming an even number of copies of $\rho$.  An even number of applications is important because one application of $\mathcal{M}$ will pass local information from $\rho$ onto $\sigma$, but with two applications of  $\mathcal{M}$ this local information cancels out.  Since pumped-Zinf creates an imperfect parity projection, it is possible to entangle qubits $A$ and $B$ of $\sigma$ without destroying prior entanglement held with third parties.  We illustrate that $\sigma$ may hold third party entanglement by lines emerging from $\sigma$, and this has a natural interpretation in terms of graph state growth.  Unlike in Fig.~\ref{fig:LoMMZinf} we do not begin by preparing a locally filtered $\rho_{\mathrm{Zinf}}$, and hence we eliminate the need for an additional ancillary qubit that would be required to perform the POVM  measurement.}
 \label{fig:Zinfpump}
\end{figure*}

However, this extra qubit is not essential.  Instead of using an ancilla to attempt the local filter, the same resulting quantum operation can be achieved by a modified protocol.  Our modified protocol uses only local unitaries and computational basis measurements and so requires no greater technological sophistication than prior protocols.  For completeness, we will also modify our protocol to distill a high fidelity parity projection, so it can be used to either produce Bell pairs or grow entangled resources for MBQC.  We describe an entanglement pumping version of Zinf where each round involves a copy of $\rho$, and some other state $\sigma$.  As before $\rho$ can be related to a state,  $\rho_{\mathrm{Zinf}}$, by local filters $g_{A} \otimes g_{B}$.  Later we will use that these local filters can be expressed as
\begin{eqnarray}
\label{eqn:filter_in_terms_of_gamma}
	g_{A} & = & \alpha_{0} \kb{+}{A_{0}}_{A} + \alpha_{1} \kb{-}{A_{1}}_{A},  \\ \nonumber
	g_{B} & = & \beta_{0} \kb{+}{B_{0}}_{B} + \beta_{1} \kb{-}{B_{1}}_{B},
\end{eqnarray} 
where $\ket{A_{i}}$ are normalized, but not necessarily orthogonal, and similarly for  $\ket{B_{i}}$. The variables $\alpha_{i}$ and $\beta_{i}$ are complex numbers satisfying $|\alpha_{i}|, |\beta_{i}| \leq 1$.  Our pumped-Zinf protocol requires no additional ancilla qubits to perform local filtering, and is capable of distilling probabilistic parity projections for graph state growth (see Fig.~\ref{fig:Zinfpump} for outline).  We use $A1$ and $B1$ to denote qubits holding one copy of $\rho$, and $A2$ and $B2$ as the qubits in the state $\sigma$ that we wish to parity project.  Our pumped-Zinf protocol is then:
\begin{enumerate}
  \item    Prepare qubits $A1$ and $B1$ in the base pair state $\rho$;
  \item    Perform local unitaries  $U_{A} \otimes U_{B} $ ;
  \item    Measure qubits $A1$ and $B1$ in the computational basis, post-selecting on the outcome $\ket{1}_{A1}\ket{1}_{B1}$;
  \item   Perform local unitaries $V_{A} \otimes V_{B} $ ;
  \item    Measure qubits $A1$ and $B1$ in the computational basis, post-selecting on the outcome $\ket{1}_{A1}\ket{1}_{B1}$;
  \item    Repeat for subsequent rounds an \textit{even} number of times;
\end{enumerate}
The protocol is quite complex and involves a number of new unitaries:
\begin{eqnarray}
U_{A} & = & \left( \kb{1}{A_{0}} + \kb{0}{\bar{A}_{0}} \right)_{A1}  \otimes \kb{0}{0}_{A2}   +  \left( \kb{1}{A_{1}} + \kb{0}{\bar{A}_{1}} \right)_{A1}  \otimes \kb{1}{1}_{A2} ,  \\ \nonumber
U_{B} & = & \left( \kb{1}{B_{0}} + \kb{0}{\bar{B}_{0}} \right)_{B1}  \otimes \kb{0}{0}_{B2}  +  \left( \kb{1}{B_{1}} + \kb{0}{\bar{B}_{1}} \right)_{B1}  \otimes \kb{1}{1}_{B2} , \\ \nonumber
V_{A} & = &  W(\alpha_{0})_{A1} \otimes \kb{1}{0}_{A2} + W(\alpha_{1})_{A1} \otimes \kb{0}{1}_{A2}, \\ \nonumber
V_{B} & = &  W(\beta_{0})_{B1} \otimes \kb{1}{0}_{B2} + W(\beta_{1})_{B1} \otimes \kb{0}{1}_{B2} ,
\end{eqnarray}						
where $W( c )$ can be any one qubit unitary satisfying $\bra{1}W(c)\ket{1}=c$.  The barred kets, e.g. $\ket{\bar{A}_{0}}$, are orthogonal to their unbarred counterpart so as to ensure unitarity.  The role of these unitaries in pumped-Zinf can appear opaque at first, and so we pause to relate them to something more familiar.  Consider local operations at location A for steps 2-5 and we find:
\begin{eqnarray}
\label{eqn:equilvalence}
	\bra{1}_{A1}  V_{A} \kb{1}{1}_{A1} U_{A}	& = &  \alpha_{0} \bra{A_{0}}_{A1} \otimes \kb{1}{0}_{A2}  + \alpha_{1} \bra{A_{1}}_{A1} \otimes \kb{0}{1}_{A2} , \\ \nonumber
	& = & X_{A2}  \bra{1} CX_{A1}^{A2} g_{A1}  ,
\end{eqnarray}
and so these local operations emulate the behavior of performing the desired local filter, followed by a control-not and post-selected measurement.  Hence, this involved protocol manages to reproduce the effect of local filtering and distillation without the use of an additional ancillary qubit!  Actually, a local filter is performed, via the POVM measurement in steps $4-5$, which reuses qubits $A1$ and $B1$ after they where measured in step $3$.  Our protocol can be thought of as a temporal reordering where local filtering is performed after distillation, rather than before, hence saving valuable spatial resources. 

A single successful round of the protocol generates a quantum operation, which we denote $\mathcal{M}(\sigma)$.  In each application of  $\mathcal{M}(\sigma)$ the presence of local information has the effect of partially projecting $\sigma$ into some separable state.  However, the preferred subspace changes for alternate rounds, cancelling itself out after an even number of applications of  $\mathcal{M}$.  This cancelling of local information emerges from continual bit-flipping, which is explicitly presented on the R.H.S of Eq.~(\ref{eqn:equilvalence}).  If these bit-flips where not present then local information would accumulate.   Hence, we always apply our protocol for an even number of rounds, and actually work with $\mathcal{E}(\sigma)=\mathcal{M}(\mathcal{M}(\sigma))$, where:
\begin{eqnarray}
	\mathcal{E}(\sigma) & \propto &  \mu_{+} P^{+} \left(  (1- q_{+}) \sigma + q_{+}  Z_{A2} \sigma  Z_{A2}   \right)P^{+} \\ \nonumber
	&& + \mu_{-} P^{-} \left(  (1- q_{-}) \sigma + q_{-}  Z_{A2} \sigma  Z_{A2}   \right)P^{-} ,
\end{eqnarray}
where $P^{+}$ ($P^{-}$) are even (odd) parity projectors, $P^{\pm}=(\unity \pm Z_{A2}Z_{B2})/2$. As with the original Zinf protocol, this is related to the concurrence eigenvalues or $\rho$, such that:
\begin{eqnarray}
	\mu_{+}  = ( \lambda_{1}+\lambda_{4} )^{2} & , & q_{+} = 2 \lambda_{1}\lambda_{4} / \mu_{+} , \\ \nonumber
	\mu_{-}  = ( \lambda_{2}+\lambda_{3} )^{2} & , & q_{-} = 2 \lambda_{2}\lambda_{3} / \mu_{-} .
\end{eqnarray}
Expressing the quantum operation in terms of these variables proves useful when calculating the effect of repeated pumping.  It is helpful to compare these results with the analogous Eq.~(\ref{Zinfresult}), where the terms match up\footnote{To see this one has to multiply out the new variables, such as $\mu_{+}(1-q_{+})=\lambda_{1}^{2}+\lambda_{4}^{2}$, to see they equate to the earlier expressions.}.  We can also calculate the  Jamiolkowski fidelity\cite{GLN01a} of $\mathcal{E}$ compared to a perfect parity projection $P^{+}$, and one finds that one application of $\mathcal{E}$ has the same fidelity as one round of the unmodified Zinf protocol  (see Eq.~(\ref{OneRoundFid})).

Now consider $2n$ successful rounds of pumped-Zinf, which is equivalent to $n$ applications of the quantum operation $\mathcal{E}$:
\begin{eqnarray}
	\mathcal{E}^{n}(\sigma) & \propto &  \mu_{+}^{n} P^{+}  \mathcal{D}_{+}^{n}( \sigma )  P^{+}  + \mu_{-}^{n} P^{-}  \mathcal{D}^{n}_{-}( \sigma ) P^{-} ,
\end{eqnarray}
where $\mathcal{D}^{n}_{\pm}(\sigma)$ represents $n$ applications of a dephasing channel:
\begin{equation}
 	\mathcal{D}_{\pm}(\sigma)  =  (1-q_{\pm}) \sigma + q_{\pm}Z_{A2} \sigma Z_{A2}.
\end{equation}
Examining the quantum channel $\mathcal{E}^{n}(\sigma)$ we can see two separate effects at work: (a) as $n$ increases the odd parity projection component will vanish (assuming $\mu_{+}>\mu_{-}$); (b) the quantum operation becomes more dephasing.   In fact the Jamiolkowski fidelity of $\mathcal{E}^{n}$ factorizes with respect to these separate effects:
\begin{equation}
	F = \frac{\mu_{+}^{n}}{\mu_{+}^{n}+\mu_{-}^{n}}  \left( \sum_{k: \mathrm{odd}}^{n} {n \choose k} (1-q_{+})^{n - k} q_{+}^{k} \right) .
\end{equation}
Such behavior is not unique to our protocol but occurs in other pumped protocols\cite{C01a,JTSL02a}.  Earlier we commented that some protocols only require a single pumping level to distill certain Bell diagonal states to unit fidelity.  These special, easily distillable, states need to have only one vanishingly small eigenvalue.  Here we have a much stronger result:  if a base pair $\rho$ has a vanishingly small concurrence eigenvalue $\lambda_{4} = 0$, then it can be distilled to unit fidelity with only a single pumping level.  We can distill towards unit fidelity because there is no dephasing in the even parity subspace, since $\lambda_{4}=0$ entails $q_{+}=0$, which gives the simple fidelity expression:
\begin{equation}
	F ( \lambda_{4}=0)= \frac{\mu_{+}^{n}}{\mu_{+}^{n}+\mu_{-}^{n}} .
\end{equation}
This result substantially broadens the class of mixed states that can enable quantum computing with only two qubits per location.  

What if a base pair does not have a small enough concurrence eigenvalue? In this instance it seems unavoidable that an extra pumping level is required to enable high fidelity distillation.  When pumping Zinf with two levels, we prepare $\sigma$ in the state $\kb{+,+}{+,+}$ so that  $\mathcal{E}^{n}(\sigma)$ is a LoMM mixed state.  This LoMM mixed state is then used in the next level of pumping with no local filtering required.

\section{Conclusions}

We have proposed two protocols,  LoMM and Zinf, for entanglement distillation.  Our protocols can achieve high fidelities in fewer rounds of distillation than previous protocols.  The mixed states we produce have fidelities that are a function of the concurrence eigenvalues encountered when calculating the concurrence of the base pair.  Our protocols achieve this by exploitation of local information, which is discarded in DEJMPS.  Although LoMM can attain high fidelities in a single round,  in the strong local information regime it suffers from a vanishing success probability, and hence yield.  Whereas Zinf can achieve the same fidelity for only a small overhead in success probability.  Consequentially, symmetrically scheduled Zinf achieves much higher yields for base pairs with large amounts of local information.

We have also shown that Zinf can be modified into a compact entanglement pumping protocol.  Furthermore, pumped-Zinf enables measurement-based quantum computing in distributed architectures with a very small number of qubits per local site, for any base pair.  In general pumped-Zinf requires 3 qubits per local site. However, only two qubits per location are needed when the base pair has a vanishing concurrence eigenvalue.  

Zinf offers increased yields when spatial resources are abundant and can function when spatial resources are very constrained.  Since Zinf has much to offer on both extremes of the spatial scale, we also expect that it will have applications in the intermediate regimes. 

I would like to thank Dr Simon Benjamin and Erik Gauger for conversations relating to an ancestor of this paper, in which only the LoMM protocol was considered.   This work has been funded by the Royal Commission for the Exhibition of 1851, and was supported by the National Research Foundation and Ministry of Education, Singapore.

\end{document}